\documentclass[prb,twocolumn,showpacs,preprintnumbers,amsmath,amssymb]{revtex4}
\usepackage{graphicx}
\usepackage{bm}
\begin{document}

\title{Electron-hole pair creation by atoms incident on a metal surface}

\author{J.R. Trail}
\email{jrt32@cam.ac.uk}
\affiliation{Department of Physics, University of Bath, Bath BA2 7AY, UK}
\author{D.M. Bird}
\affiliation{Department of Physics, University of Bath, Bath BA2 7AY, UK}
\author{M. Persson}
\affiliation{Department of Applied Physics, Chalmers/G\"oteborg University, 
S-412 96 G\"oteborg, Sweden}
\author{S. Holloway}
\affiliation{Surface Science Research Centre, University of Liverpool, 
Liverpool L69 3BX, UK}

\date{March, 2003}

\begin{abstract}
Electron-hole pair creation by an adsorbate incident on a metal
surface is described using \textit{ab initio} methods.  The approach
starts with standard first principles electronic structure theory, and
proceeds to combine classical, quantum oscillator and time dependent
density functional methods to provide a consistent description of the
non-adiabatic energy transfer from adsorbate to substrate.  Of
particular interest is the conservation of the total energy at each
level of approximation, and the importance of a spin transition
as a function of the adsorbate/surface separation.  Results are
presented and discussed for H and D atoms incident on the Cu(111)
surface.
\end{abstract}


\maketitle

Understanding the fundamental processes involved in gas-surface
interactions is an important goal, from both a pure and applied
scientific perspective.  In recent years much progress has been made
in understanding these interactions, \cite{darling95} but a key
phenomenon that has received little attention is energy dissipation
into substrate degrees of freedom for an adsorbate incident upon a
surface - an essentially non-adiabatic process.  Energy loss by phonon
excitation has recently been described successfully (notably by Wang
\textit{et al.} \cite{wang01} and Busnengo \textit{et
al}. \cite{busnengo01} ), but a description of the energy loss by the
excitation of low-energy electrons at metal surfaces has not been
achieved for `real' systems.  This energy transfer mechanism is known
to be of central importance in many situations. \cite{darling95} Here
we report \textit{ab initio} calculations of electron-hole pair
creation for H/Cu(111).

In a previous paper \cite{trail01} we have described a method to
calculate the classical force experienced by an atom incident on a
metal surface due to excitation of electrons in the system - a nearly
adiabatic process in which many low-energy electron-hole pairs are
excited.  This method is based upon Time Dependent Density Functional
Theory (TDDFT) together with a nearly-adiabatic approximation that allows 
the definition of a position-dependent friction coefficient.  In
Ref. \onlinecite{trail01} we principally described the development of
the theory and parallel algorithm necessary to obtain this friction
coefficient.  Preliminary results were presented for H on Cu(111), and
discussed only in terms of convergence behaviour and basic viability
of the method.

In a recent Letter \cite{trail02} we have briefly described a further
extension of this work.  The creation of electron-hole pairs due to
the adsorbate/surface interaction is described as the excitation of
quantum oscillators by a driving force - the well known Forced
Oscillator Model (FOM).  This approach allows the prediction of the
energy spectrum of excited electrons.  The results obtained may be
compared to the recent experimental work of Nienhaus and co-workers,
who have directly measured the hot electrons and holes created at
metal surfaces by the adsorption of thermal hydrogen and deuterium
atoms in the form of a `chemicurrent' in a Schottky
diode. \cite{nienhaus99,nienhaus00,nienhaus02,gergen01} Good agreement
was found between predicted and experimental results for H/Cu(111).

Our aim in this paper is to provide a more complete description of the
theory and approximations employed in these calculations, and to show
explicitly the relationship between the classical description of the
adsorbate dynamics and the quantum description of the excited
electrons.  A further important goal is to define clearly the
approximations in the theory and their physical consequences.  Of
particular interest is the spin transition (from spin-polarised
to non-magnetic) that can occur when an open-shell adsorbate approaches a
metallic surface. This spin transition is found for the H/Cu system
and in this case a direct application of the friction coefficient and
FOM results in an unphysical singularity in the rate of energy transfer.  
In Ref. \onlinecite{trail02} we provided a brief argument as to how this
problem may be resolved.  Here we present a more complete discussion of
this singular behaviour, together with further details of the results 
obtained using the FOM for the properties of the excited electron-hole 
pairs. In a recent paper Gadzuk \cite{gadzuk02} 
has also analysed the excitation and detection of chemicurrents. He uses 
a three step model consisting of (i) the excitation of hot electrons, 
(ii) their transport across the thin metal film of the Schottky diode, and 
(iii) their transmission across the metal-semiconductor interface. Our
principal aim in this paper is to provide a detailed theory of step (i)
that goes beyond the simple model expressions used by Gadzuk.

The approach we use may be naturally divided into three stages.  A
standard first principles DFT calculation of the adsorbate/surface
system is carried out to obtain the Kohn-Sham (KS) states and
self-consistent potential for the adsorbate at various positions.  A
friction coefficient is then obtained from the KS states and potential
which gives the energy loss of the incident adsorbate due to
electron-hole pair creation.  This friction coefficient provides a
description of the average non-adiabatic energy transfer behaviour of
the system, and is derived by applying TDDFT and a nearly-adiabatic
approximation.  We may then solve a classical equation of motion to
obtain a trajectory for the incident adsorbate.  Finally the response
of the electron gas to the time-varying potential due to this
trajectory is found by applying the FOM to obtain a semi-classical
(classical atomic motion coupled to quantum metallic electrons)
description of electron-hole pair creation.  In the next section the
formalism that leads to the friction coefficient is developed, and 
results given for H/Cu(111). In section \ref{sec:fom} the FOM is 
presented, together with the approximations used in our implementation, 
and discussed.  In section \ref{sec:res} we show how this method is used 
to predict a variety of experimentally measurable quantities. Results 
for H and D atoms incident on the Cu(111) surface are presented and 
analysed.

\section{Friction Coefficient}
\label{sec:fric}
We begin with a description of the classical motion of an adsorbate
coupled to the electronic states of the substrate.  An
approach is required that extends beyond the Born-Oppenheimer
approximation of nuclear motion on a single potential-energy surface
to include non-adiabatic energy loss of the incident nuclei to the
substrate electrons.  The nuclei of the substrate are considered
stationary, although our approach can easily be generalised to include
their motion.  Our goal in this section is to outline the theory used
to obtain a friction description of the energy transfer (as first 
derived by d'Agliano \textit{et al.} \cite{dagliano} and Blandin 
\textit{et al.} \cite{blandin} and used by Hellsing and 
Persson \cite{hellsing84} for damping of vibrational modes; see 
also Head-Gordon and 
Tully \cite{headgordon95} and Plihal and Langreth \cite{plihal98}) 
and apply this to the H/Cu(111) system. Results for
the friction coefficient are found to exhibit singular behaviour. We
explain the source of this unphysical singularity and describe an
approach taken to remove it.

The time-dependent motion of an adsorbate results in a non-adiabatic
energy transfer to the many-electron system of the substrate, which 
can be handled in the nearly adiabatic approximation. To ease the notation 
we restrict the discussion to a single atomic nucleus following a trajectory
$z(t)$. In general the non-adiabatic energy transfer from the nucleus to the
many-electron system can be expressed as (see Appendix),
\begin{equation}
\dot{E}_{\textrm{non-ad}}(t) = 
\dot{z}(t)\int d\mathbf{r}\frac{dV_{\textrm{ext}}(\mathbf{r},z(t))}{dz}
                                        \delta n(\mathbf{r},t).
\label{eq:nonad}
\end{equation} 
Here, $V_{\textrm{ext}}(\mathbf{r},z)$ is the electron-nucleus interaction
potential and $\delta n(\mathbf{r},t) = n(\mathbf{r},t) -
n_0(\mathbf{r},z)$ is the deviation of the instantaneous electron density
$n(\mathbf{r},t)$ from the ground state electron density $n_0(\mathbf{r},z)$ 
when the nucleus is fixed at position $z(t)$. In our case, the time-dependent 
motion of the nucleus is so slow that $\delta n(\mathbf{r},t)$ is expected 
to be small and can be handled in a nearly adiabatic approximation. In this 
approximation $\delta n(\mathbf{r},t)$ is obtained by using the linear 
response of the electron system for a fixed position of the atomic nucleus
to the perturbation  
\begin{equation}
V_{\textrm{ext}}(\mathbf{r},z(t^\prime))- V_{\textrm{ext}}(\mathbf{r},z(t)) 
\simeq \frac{dV_{\textrm{ext}}(\mathbf{r},z(t))}{dz}(z(t^\prime)- z(t)) \ ,
\label{eq:linear}
\end{equation}
where in the last step we have assumed that the nucleus only moves a
small distance during the electronic response time. The linear response of 
the many-electron system to an arbitrary perturbation 
$\delta V(\mathbf{r},t)$ is given by
\begin{equation}
\delta n(\mathbf{r},t)=
\int_{-\infty}^{t} dt^\prime   \int d\mathbf{r}^\prime
\chi(\mathbf{r},\mathbf{r}^\prime,t-t^\prime;z) 
\delta V(\mathbf{r}^\prime,t^\prime)
\label{eq:den-den}
\end{equation}
where $\chi(\mathbf{r},\mathbf{r}^\prime,t-t^\prime;z)$ is the
density-density response kernel for the nucleus at position $z$.
Equation~(\ref{eq:den-den}) together with the perturbation of 
Eq.~(\ref{eq:linear}) result in the following expression for the 
non-adiabatic energy transfer,
\begin{equation}
\dot{E}_{\textrm{non-ad}}(t) = 
\dot{z}(t)\int_{-\infty}^t dt^\prime \Lambda(t-t^\prime;z(t))z(t^\prime).
\label{eq:EnonadRes}
\end{equation}
Here the memory function, $\Lambda$, is defined as 
\begin{widetext}
\begin{equation}
\Lambda(\tau;z) = \int d\mathbf{r}\int d\mathbf{r}^\prime
\frac{dV_{\textrm{ext}}(\mathbf{r},z)}{dz}
\left (\chi(\mathbf{r},\mathbf{r}^\prime,\tau;z) - \delta(\tau)
\int_0^\infty d\tau^\prime \chi(\mathbf{r},\mathbf{r}^\prime,\tau^\prime;z)
\right ) \frac{dV_{\textrm{ext}}(\mathbf{r}^\prime,z)}{dz} \ .
\label{eq:LambdaRes}
\end{equation}
\end{widetext}
Note that this formalism can also be applied to vibrational damping of an
adsorbate by excitation of electron-hole pairs. In this case, the expansion 
is simply around the equilibrium position.\cite{hellsing84}

The friction force description of the lossy response of the
many-electron system to the adsorbate motion is now obtained by taking
the slow adsorbate limit. In this case, we can use the low frequency limit 
of the response of the electron gas and one finds, in close analogy with 
the derivation of the vibrational damping rate,\cite{hellsing84}
\begin{equation}
\Lambda(\omega;z)= - i \eta(z) \omega + {\cal O}(\omega^2),
\label{eq:LambdaExp}
\end{equation}
where $\eta(z)$ is real.  This low frequency approximation for
$\Lambda(\omega;z)$, inserted in Eq.~(\ref{eq:EnonadRes}), gives directly 
a friction description for the energy loss
\begin{equation}
\dot{E}_{\textrm{non-ad}}(t) = \eta(z(t))\dot{z}(t)^2\ .
\label{eq:linearres}
\end{equation}
That this result gives rise to a friction force is simply understood by 
imposing energy conservation for the combined many-electron and nucleus 
system. The time derivative of the total energy then gives
\begin{equation}
M \ddot{z} = -\frac{dV_0}{dz} - \eta(z(t)) \dot{z}
\label{e2.3}
\end{equation}
where $M$ is the mass of the atomic nucleus and $V_0(z)$ is the ground
state energy of the electronic system when the nucleus is at a fixed
position $z$. In Eq.~(\ref{eq:linearres}) it can be seen that the
`memory' of the lossy response of the electronic system is removed,
hence the limit leading to Eqs.~(\ref{eq:linearres}) and (\ref{e2.3}) 
can be regarded as a Markov approximation. For a complete equation of 
motion for the adsorbate one
needs to add a stochastic force to Eq.~(\ref{e2.3}) to ensure that
the particle reaches thermal equilibrium. However, in our case the 
temperature is low and we can neglect this force.

An explicit expression for the friction coefficient can now be obtained
using time-dependent density functional theory, again in close analogy with
the derivation of the vibrational damping rate.\cite{hellsing84} In TDDFT 
the response kernel at a frequency $\omega$ is approximated by the response 
kernel for the non-interacting KS electrons and the external field is
replaced by an effective field $V_{eff}(\mathbf{r},\omega)$ which includes 
the Hartree and exchange-correlation fields. Under the widely used assumption 
of an adiabatic exchange-correlation potential, the imaginary part of the
frequency dependent memory function in Eq.~(\ref{eq:LambdaRes}) is given by
\cite{hellsing84}
\begin{widetext}
\begin{equation}
\mathrm{Im}\ \Lambda(\omega;z) = - 2 \pi \sum_{i,j}
\left| \langle \psi_i | \frac{d V_{eff}(\omega)}{dz} |\psi_j \rangle 
\right| ^2 \left( f(\epsilon_i) - f(\epsilon_j) \right) 
\delta( \hbar\omega-\epsilon_i + \epsilon_j )
\label{e2.8}
\end{equation}
\end{widetext}
where $\psi_i$ and $\epsilon_i$ are the KS eigenstates of the electrons at
position $z$ of the adsorbate, and $f(\epsilon)$ is the Fermi-Dirac occupation 
function. The friction coefficient is now, according to Eq.~(\ref{eq:LambdaExp}), 
obtained from the low frequency limit of $\mathrm{Im}\ \Lambda(\omega)$. By 
making a straightforward generalisation to spin-polarised TDDFT, and changing 
our notation slightly, we obtain
\cite{trail01}
\begin{equation}
\eta(z)= \pi \hbar
         \sum_{\sigma}
         \sum_{\alpha,\alpha'} \left|
         \langle \epsilon_{\mathrm F} \alpha,\sigma | 
          \frac{d V_{\sigma}}{dz} | 
          \epsilon_{\mathrm F} \alpha',\sigma \rangle
         \right|^2
\label{e2.9}
\end{equation}
where $\alpha$ and $\alpha^\prime$ are supplementary quantum numbers 
for states on the Fermi surface, and $V_{\sigma}(\mathbf{r},z)$ is the
KS potential for spin $\sigma$. It is important to note that the static 
KS potential, $V_{\sigma}(\mathbf{r},z)$, for spin $\sigma$ appears in
Eq.~(\ref{e2.9}) as a result of the low frequency limit and not due to
an additional approximation or assumption, hence this is the correct
result within TDDFT for an adiabatic XC functional.

The results in Eqs.~(\ref{e2.3}) and (\ref{e2.9}) allow us to
calculate the classical trajectory of an adsorbate interacting with a
substrate, using a friction coefficient to describe the non-adiabatic
energy loss to the substrate electrons.  This description is valid
within TDDFT (of course an approximate XC functional must be chosen
for calculations) in combination with two approximations.  First, we
assume the substrate electrons respond nearly-adiabatically to the
motion of the adsorbate in the sense that the instantaneous electron
density is always close to the ground state density. The second, and
most significant, approximation is the assumption that the adsorbate
moves so slowly that the Markov limit can be taken to remove the memory
of the lossy response of the electronic system. The validity of this
approximation depends on the assumption of a linear frequency
dependence of the memory function over the frequency range set by the
time-dependence of the adsorbate motion. Such a frequency dependence
is expected over the range where the density of states for
electron-hole pairs is essentially linear. In the case of a jellium
surface the corresponding energy range is several eV and the Markov
limit should be a good approximation for atoms with a kinetic energy
of a few eV.\cite{hellsing84} However, as discussed in more detail below, 
we find that the adsorbate motion through the spin transition is 
not slow, and in this case the Markov approximation fails. 

\subsection{H/Cu(111): Spin transition and singularities}
We begin with a standard, self-consistent plane-wave pseudopotential
calculation of the required KS states and potentials, for H atoms
moving perpendicular to the surface above the top site of Cu(111).  To
describe the surface a five-layer slab together with a vacuum gap
equivalent to another five empty layers is employed.  Calculations are
performed using a $2\times2$ in-plane super-cell, XC effects are
described by a spin-polarised version of the PW91 functional,
\cite{perdew96} a Troullier-Martins \cite{troullier91} pseudopotential
is used for Cu, and H is represented by a Coulomb potential.  A
plane-wave cut-off of 830 eV is used and 54 k-points are included in
the full surface Brillouin zone, together with a Fermi surface
broadening of 0.25 eV.  Total energy calculations are performed for a
range of heights between 1.0 and 4.0 \AA\ above the surface, and for
heights $\pm h$ around these points from which the deformation
potential, $dV_{\sigma}/dz$, is calculated using
a finite difference.  For the results given here, $h=0.02$ \AA.  Full
details of the calculation are given in Trail \textit{et
al}. \cite{trail01} Friction coefficients are evaluated for the same
range of heights and for motion perpendicular to the surface.  Care
must be taken to interpret the super-cell geometry correctly when
evaluating Eq.~(\ref{e2.9}) within a plane-wave basis since we are
interested in the isolated motion of an atom above the surface, not
the coherent motion of a mono-layer.  It is also important to perform
a correct discretisation of Eq.~(\ref{e2.9}) within the finite
available sampling of $\mathbf{k}$ space to obtain an accurate
friction coefficient. \cite{trail01} In keeping with the natural
application of spin-polarised DFT, we begin by assuming
spin-adiabaticity.  This means that the total energy is minimised with
respect to the magnetisation density as well as the charge density.
We refer to this in what follows as the `free spin' case.

As a preliminary test it is encouraging that the friction coefficient
at the total energy minimum predicts a lifetime of $0.8$ ps for the
perpendicular vibrational mode of atomic hydrogen on Cu(111), a value
that compares well with the $0.7$ ps deduced from the results of
Infrared Reflection Absorption Spectroscopy experiments.\cite{lamont95} 
Here we have assumed that the amplitude of the mode is small enough 
that no significant variation of $\eta(z)$ occurs, and that the 
potential is harmonic.

\begin{figure}[t]
\includegraphics[scale=0.95]{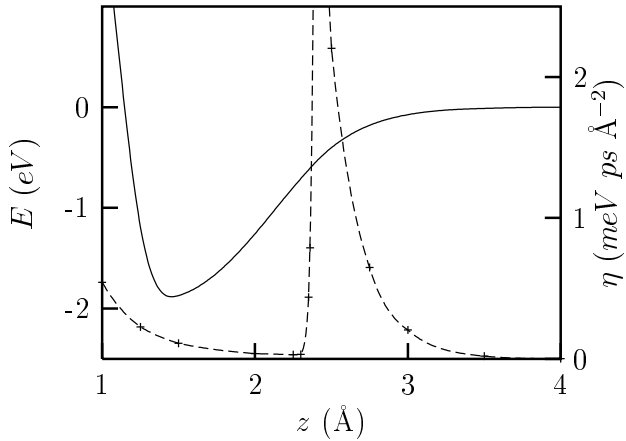}
\caption{\label{fig1}
Potential energy (solid line) and friction coefficient (dashed line)
for H atom in perpendicular motion above the top site of Cu(111). }
\end{figure}

\begin{figure*}[t]
\includegraphics[scale=0.95]{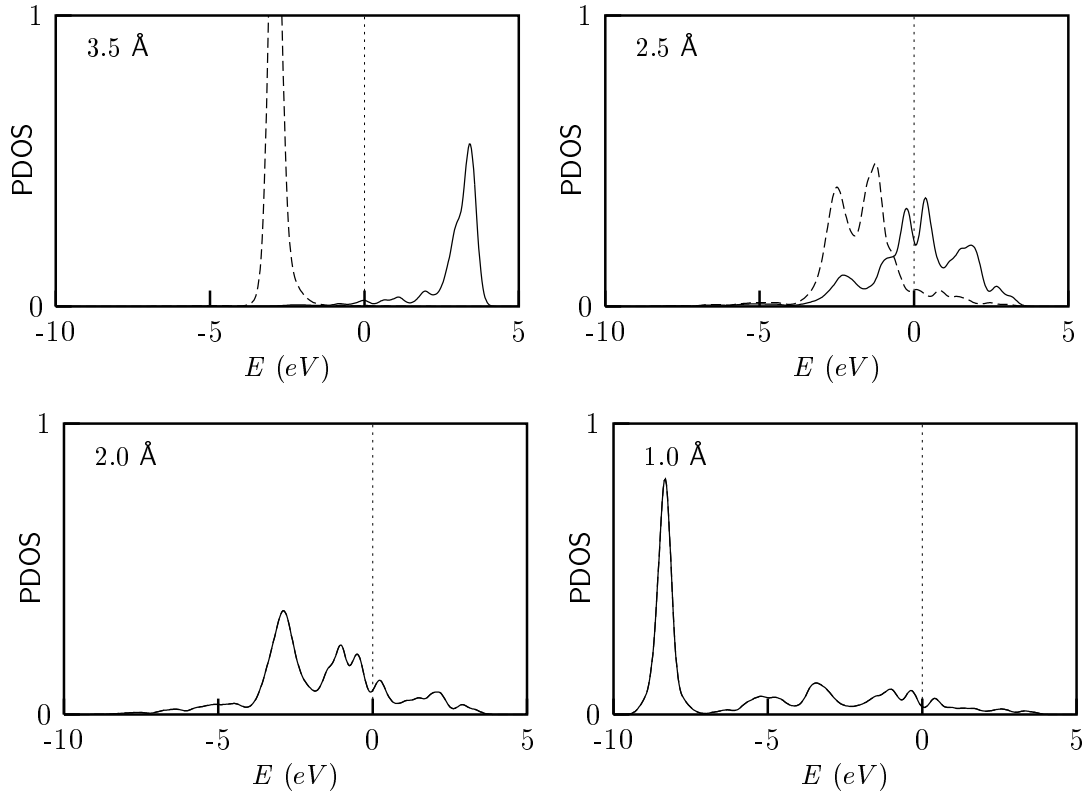}
\caption{\label{figdos}
Densities of states projected onto the hydrogen $1s$ orbital at 
different heights above the Cu(111) surface. Dashed and solid lines 
show results for majority and minority spins respectively. Fermi
energy is at 0 eV.}
\end{figure*}

Figure\ \ref{fig1} shows the friction coefficient and potential energy
for the range of heights considered.  It is immediately apparent that
for H close to the surface and very far from the surface our results
may be realistic, but that as the atom approaches $z_0=2.39$ \AA\ from 
above
the friction coefficient shows singular behaviour.  Analysis of these
results shows that close to $z_0$, $\eta(z) \sim (z-z_0)^{-1}$.
We also find that $z_0$ is the height at which the system makes a
transition from a spin-polarised ($z > z_0$) to spin-degenerate ($z
\leq z_0$) state. The changes in the
electronic structure that accompany the spin transition are shown by
the density of states projected onto the hydrogen 1$s$ orbital
(Fig.~\ref{figdos}). When the atom is well separated from the surface
the density of states shows narrow resonances with the majority spin
state fully occupied and the minority spin state empty. As the atom
approaches the surface (see figure for 2.5\ \AA) both spin states
broaden and begin to merge. At a height of 2.0\ \AA\ the two spin 
states have become degenerate and there is no net polarisation. For 
closer atom-surface separations the H-related states continue to 
drop in energy.

The results of Fig.~\ref{fig1} clearly show that the singular
behaviour of the friction coefficient is a consequence of the spin
transition.  The effect of this singularity in $\eta$ may be
assessed by considering the energy loss, $\Delta E$, for an adsorbate
following an arbitrary trajectory $z(t)$.  For $\eta(z) = \eta_0
z^{-1}$
\begin{eqnarray}
\Delta E =& \int_{-\infty}^{\infty} \eta(z) v^2 dt   \\ \nonumber 
         =& \eta_0 \int_{z_i}^{z_f} v {z'}^{-1} dz',
\label{e2.10}
\end{eqnarray}
where $v=\dot{z}$, $z_i$ is the initial height and $z_f$ is the final
height.  For any real trajectory, $\Delta E$ must be finite so either
the trajectory does not reach the singularity (at $z=0$) for any time,
or at $z=0$ the velocity is zero.  In other words, any real trajectory
stops at or above the singularity, no matter how fast the adsorbate
impinges on the surface.  For example, a flat potential results in a
trajectory that starts at $z_i$ with a velocity of $-v_i$, and stops
at $z_{f}=z_{i} \exp ({-\frac{mv_{i}}{\eta_{0}}} )$.  This `infinite
stopping power' is obviously unphysical, and the source of this
behaviour must be identified.

We begin by showing that the singular behaviour is an expected
consequence of the theory applied here, and is not a numerical
artifact.  Equation (\ref{e2.9}) contains the expression
$dV_{\sigma}/dz$.  In terms of the spin up and
spin down charge densities, $\rho_\uparrow$ and $\rho_\downarrow$, we
define the total charge density, $\rho= \rho_\uparrow +
\rho_\downarrow$, the total spin $S=\int (\rho_\uparrow -
\rho_\downarrow) d^3 {\mathbf r}$ and the relative distribution of
magnetisation $\zeta= ( \rho_\uparrow - \rho_\downarrow )/S$.
Expressing the charge and magnetisation densities in terms of these
three quantities allows us to separate out the gross variation due to
changes in total spin by giving the variation of the potential with
$z$ as
\begin{widetext}
\begin{equation}
     \frac{dV_{\sigma}({\mathbf r })}{dz} =
     \int \left(
     \frac{\delta      V_{\sigma}({\mathbf r })}{\delta \rho({\mathbf r'})}
     \frac{\partial \rho({\mathbf r'})}{\partial z} +
     \frac{\delta      V_{\sigma}({\mathbf r })}{\delta \zeta({\mathbf r'})} 
     \frac{\partial    \zeta({\mathbf r'})}{\partial z}
     \right) d^3 {\mathbf r'} +
     \frac{\partial    V_{\sigma}({\mathbf r })}{\partial  S}
     \frac{\partial                  S}{\partial z}.
\label{e2.11}
\end{equation}
\end{widetext}
We concentrate on the third term of this equation since this
explicitly describes the effect of changes in the total spin.  Our
calculations show that $S \sim (z-z_0)^{ \frac{1}{2} }$, as expected
for a second order phase transition within a mean field theory.
This implies that, provided $\frac{\partial V_{\sigma}({\mathbf r
})}{\partial S}$ has a constant part to its variation with $S$ as $S
\rightarrow 0$ (and there is no reason to believe it has not), the
resulting $(z-z_0)^{ -\frac{1}{2} }$ behaviour will dominate the
deformation potential near $z_0$.  Through the definition of the
friction coefficient in Eq.~(\ref{e2.9}) this results in the $\sim
z^{-1}$ behaviour of $\eta$ above the transition point.

Given that the singularity arises naturally in our friction
coefficient description of electron-hole pair creation, it is
important to ask which of the approximations in the underlying theory
breaks down in the vicinity of the spin transition.  As
discussed above, a major assumption is that the linear part of the 
expansion of $\Lambda(\omega;z)$ in Eq.~\ref{eq:LambdaExp} provides
an accurate approximation for $\mathrm{Im}\ \Lambda(\omega;z)$ over the
range of $\omega$ where the trajectory is significant.  It is a
failure of this assumption that results in the singular behaviour;
any non-zero velocity of the adsorbate exactly at the spin transition
means that the nearly adiabatic approximation breaks down. The 
spin cannot relax instantaneously, as is assumed in the free spin case.
Given that some `memory' must be retained in reality, the Markov
approximation cannot provide a realistic description close to
the spin transition. The timescale for spin relaxation will in
practice be set by the rate of hopping of electrons between 
adsorbate and substrate and this in turn is governed by the
width of the projected density of states shown in Fig.~\ref{figdos}.
How can the unphysically fast spin relaxation that naturally arises
in our nearly adiabatic theory be prevented? We have chosen to
adopt the simplest solution to this problem, namely to keep the
total spin fixed for the whole trajectory of the adsorbate. 
Our justification for this derives largely from the results we
obtain. We show below that, except in the immediate vicinity of 
the transition point, the potential energy and friction coefficient
are very similar for the free spin and fixed spin cases. Also,
it is shown in section \ref{sec:res} that the peak in the
friction coefficient near $z_0$ does not provide a dominant
contribution either to the total rate of electron-hole pair
creation or to the spectrum of the electron hole pairs, provided
the singularity is smoothed. Essentially, our results show that,
provided the unphysical singularity is removed in a reasonable
way, the results for electron-hole pair creation are not strongly
dependent on the way in which the singularity is smoothed. In
another paper we address this point in more detail, by using 
the Newns-Anderson model of the adsorbate-substrate interaction
to analyse the dynamics close to the spin transition.
\cite{newpaper} 

For the free spin case Eq.~(\ref{e2.11}) can be viewed as
$\lim_{\omega \rightarrow 0}$ of the derivative of the TDDFT effective
potential, as required in Eq.~(\ref{e2.8}).  The fixed spin
approximation takes $\lim_{\omega \rightarrow 0}$ for the variation of
the charge and spin densities, but $\lim_{\omega \rightarrow \infty}$
for the variation of the \textit{total} spin, $S$.  Essentially this
corresponds to assuming that the total charge density, $\rho$, and the
relative distribution of magnetisation, $\zeta$, respond adiabatically
to the motion of the incident adsorbate, but the total spin, $S$,
responds in the opposite impulsive limit - that is, for the time
scales considered, the total spin does not have time to change.
For the fixed spin case Eq.~(\ref{e2.11}) is replaced by
\begin{equation}
   \left.  \frac{dV_{\sigma}({\mathbf r })}{dz} \right|_{S} =
     \int 
     \frac{\delta      V_{\sigma}({\mathbf r })}{\delta \rho({\mathbf r'})}
     \frac{\partial \rho({\mathbf r'})}{\partial z} +
     \frac{\delta      V_{\sigma}({\mathbf r })}{\delta    m({\mathbf r'})} 
\left.   \frac{\partial    m({\mathbf r'})}{\partial z}  \right|_S
     d^3 {\mathbf r'},
\label{e2.18}
\end{equation}
where $m({\mathbf r})= \rho_\uparrow - \rho_\downarrow $ is the
standard KS magnetisation density, and variations are constrained such
that $S$ is unchanged. \cite{fixednote}  In this form the $z^{-1}$ 
singularity due to the change in total spin is removed.

\begin{figure}[t]
\includegraphics[scale=0.95]{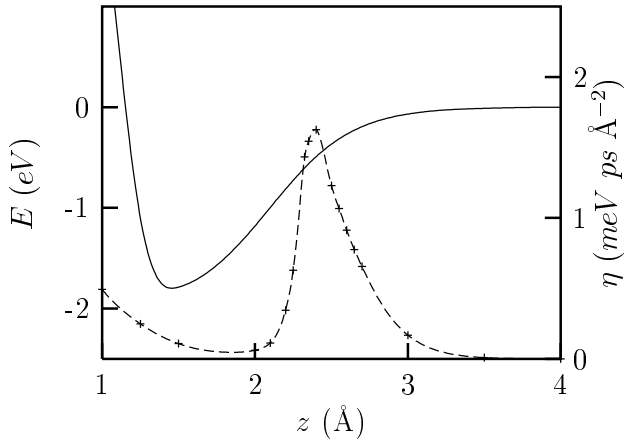}
\caption{\label{fig2}
Potential energy (solid line) and friction coefficient (dashed line)
for H atom in perpendicular motion above the top site of Cu(111).
Results are obtained using the fixed spin approximation.
}
\end{figure}

Figure\ \ref{fig2} shows the friction coefficient and potential energy
for the fixed spin approximation with $S=1$, the total spin for the H
atom far from the surface.  The potential energy agrees well with the
previous free spin results - the maximum difference occurs at the
minimum, where the fixed spin energy is $\sim 0.08$ eV greater than
that for free spin.  Agreement between the calculated $\eta(z)$ away
from the transition height ($z_0=2.39$ \AA) is also good.  Since no
singularity is present the problem of an infinite stopping power is
removed.

\section{Forced Oscillator Model}
\label{sec:fom}
In the previous section a semi-classical description of the motion of
an adsorbate was given, in the sense that the classical motion of the
adsorbate atom can be obtained by taking into account the energy
losses due to excitations of the electrons of the metal surface.  In
this section our aim is to describe in more detail the excitation of
the electron gas itself.  This will enable us to obtain the spectrum
of the excited electron-hole pairs and thus make a connection to a
variety of experimentally measurable quantities.  The basis of the
approach is the well known Forced Oscillator Model (FOM), where here
the oscillators are electron-hole pairs, and these are `forced' by the
changing potential due to the motion of the adsorbate.  Past
applications of the FOM have involved the investigation of simple
model systems for electron-hole pairs (see, for example, Sch\"onhammer
and Gunnarson
\cite{schonhammer84,schonhammer80,schonhammer81,gunnarsson82},
Minnhagen \cite{minnhagen82} and Brako and Newns \cite{brako81}) and
more detailed phonon models (Persson and Harris
\cite{persson87,persson87b}).  Our goal here is to describe a FOM
approach rooted in our \textit{ab initio} description of friction, and
assess to what extent the FOM is consistent with TDDFT.

We begin with the energy distribution function, $P(\omega)$, defined
as the probability that the electron gas, subjected to a potential
$V(t)$, is excited to energy $\hbar \omega$ above the ground state,
after the interaction has taken place. \cite{schonhammer84} By
applying a second order cumulant expansion \cite{kubo62} to the
appropriate matrix element it is possible to obtain an approximation
to $P(\omega)$ that is exact for some special cases, and is a good
approximation for slowly varying potentials of arbitrary magnitude.
Within this truncated cumulant expansion $P(\omega)$ is given by \cite{schonhammer84}
\begin{equation}
P(\omega)=\frac{1}{2\pi} \int_{-\infty}^{\infty}
          {\mathrm e}^{P_{s}(t)-\alpha_0}
          {\mathrm e}^{ {\mathrm i} \omega t} dt     
\label{e3.1}
\end{equation}
where $\alpha_0=\int_{0}^{\infty} P_s(\omega) d\omega$ is the average
number of electron-hole pairs excited, ${\mathrm e}^{-\alpha_0}$ is
the Debye-Waller factor and gives the probability of the system
remaining unexcited, and
\begin{equation}
P_s(t)= \int_{0}^{\infty} P_s(\omega) {\mathrm e}^{-{\mathrm i} \omega t } d\omega.
\label{e3.2}
\end{equation}
Equations (\ref{e3.1}) and (\ref{e3.2}) may be understood as a
multi-excitation expansion of the probability of the electron gas
being excited by $\hbar \omega$ after infinite time has passed, in
terms of the probability that $n$ electron-hole excitation events have
occurred, and the probability that $n$ electron-hole excitation events
will excite the electron gas by energy $\hbar \omega$.  Here,
$P_s(\omega)/\alpha_0$ can be interpreted as the probability that a
single electron-hole excitation event is of energy $\hbar \omega$.

For non-interacting electrons, and $\omega > 0$ (defining
$P_s(\omega)=0$ for $\omega < 0$), $P_s(\omega)$ is given by
\cite{schonhammer84}
\begin{widetext}
\begin{equation}
P_s(\omega)=\frac{1}{\hbar^2 \omega^2} \sum_{ij} 
            \left| \int_{-\infty}^{\infty}
            \langle \psi_i(t) | \dot{V} | \psi_j(t) \rangle
            dt \right|^2
            \delta(\omega-\omega_i+\omega_j)
            \left( f(\omega_i)-f(\omega_j) \right)
\label{e3.4}
\end{equation}
\end{widetext}
where $| \psi_{i}(t) \rangle$ are the time dependent one-electron
states that are initially eigenstates of the unperturbed Hamiltonian,
with eigenvalues $\omega_i$, and evolve via the full perturbed
Hamiltonian.  $\dot{V}(t)$ is the time derivative of the perturbing
potential.  As described by Sch\"onhammer and Gunnarsson
\cite{schonhammer84} this expression may be simplified by making two
assumptions.  First we assume that $V(t)$ is slowly varying, so the $|
\psi_{i}(t) \rangle$ are well approximated by the time-dependent
adiabatic states of the full Hamiltonian, ${\mathrm e}^{-{\mathrm
i}\omega_i t} | \psi_{i} \rangle_t$.  Physically, this adiabatic
approximation implies that the probability of electron-hole excitation
is independent of the number of excitation events that have occurred.
For a slow variation of $V(t)$ on a time scale $T$, only states for
which $| \hbar \omega_i - \epsilon_F |\ll \hbar/T$ contribute to
Eq.~(\ref{e3.4}).  This suggests that the matrix elements may be well
described by the values taken at the Fermi level, the second
approximation by Sch{\"o}nhammer and Gunnarsson.  With these 
assumptions, Eq.~(\ref{e3.4}) becomes
\begin{equation}
P_s(\omega)=\frac{1}{\omega}
            \sum_{\alpha,\alpha',\sigma} \left| \int_{-\infty}^{\infty}
            \langle \epsilon_{\mathrm F} \alpha , \sigma
            | \dot{V} |
            \epsilon_{\mathrm F} \alpha' , \sigma \rangle
            {\mathrm e}^{-{\mathrm i} \omega t} {\mathrm d}t \right|^2,
\label{e3.5}
\end{equation}
where $| \epsilon_{\mathrm F} \alpha , \sigma \rangle$ denotes a
stationary single particle state of spin $\sigma$ on the Fermi
surface, and a supplementary quantum number, $\alpha$, has been
introduced (as in Eq.~(\ref{e2.9})).  

So far we have not stated what the non-interacting electron system 
actually is.  We now show that, if Kohn-Sham states and the KS 
potential are used in Eq.~(\ref{e3.5}), then the rate of energy 
gain by the electron gas given by Eqs.~(\ref{e3.1}) and (\ref{e3.5}) 
is identical to the energy loss rate of the adsorbate due to friction, 
given by Eqs.~(\ref{eq:linearres}) and (\ref{e2.9}). It is
shown in the Appendix (see also Ref.~\cite{minnhagen82}) that, within
the FOM, the rate at which the electrons gain energy by the 
non-adiabatic process is 
\begin{equation}
\dot{E}_{\textrm{non-ad}}(t) = \hbar       \int_{0}^{\infty}
            \omega \frac{\partial}{\partial t} P_s(\omega,t)
            d\omega,
\label{e3.6}
\end{equation}
where the $t$ variable in $P_s(\omega,t)$ has been introduced as a new
upper limit to the integral in Eq.~(\ref{e3.4}) or (\ref{e3.5}). 
Substitution of Eq.~(\ref{e3.5}) into Eq.~(\ref{e3.6}) and integration 
over $\omega$ then gives
\begin{equation}
\dot{E}_{\textrm{non-ad}}(t)=
         \pi \hbar
         \sum_{\alpha,\alpha',\sigma} \left|
         \langle \epsilon_{\mathrm F} \alpha , \sigma
         | \dot{V} | 
         \epsilon_{\mathrm F} \alpha' , \sigma \rangle
         \right|^2=\eta(z) \dot{z}^2.
\label{e3.8}
\end{equation}
This shows that the FOM implemented with adiabatic KS states and matrix
elements taken at the Fermi energy gives the same average energy gain
as the TDDFT description within the Markov approximation.

The close connection between the TDDFT and FOM approaches to energy
transfer from adsorbate to substrate implies that the FOM can be used
to provide a theory of the excitation spectrum of the electron gas
(that is, $P(\omega)$ and $P_s(\omega)$) that is consistent with the
friction description presented in the previous
section. This connection between TDDFT and the FOM means that the
singularity associated with the spin transition will also be
present in the FOM. This is clear in Eq.~(\ref{e3.5}) where $\dot{V}$
is the derivative of the Kohn-Sham potential. For the free
spin case $\dot{V}$ will be singular at the spin transition,
resulting in unphysical behaviour.  The argument that led to
constraining the total spin to be constant, thus removing the singular
behaviour, is therefore as relevant to the FOM as to the friction 
description of the energy transfer.

Equation (\ref{e3.5}) may be evaluated by the same approach used to
obtain the friction coefficient, however this requires the storage and
interpolation of a large number of matrix elements.  Due to the
approximations already made to develop the theory to this point this
effort does not seem justifiable.  Instead, we adopt an analogue of
models used by past authors to describe Fermion excitations of this
nature. \cite{brako81,minnhagen82,schonhammer84,darling95,mullerhartmann71}
It is well established that if the eigenstates can be chosen such that
the matrix element $\langle \epsilon_{\mathrm F} \alpha , \sigma |
\dot{V} | \epsilon_{\mathrm F} \alpha' , \sigma \rangle$ that appears
in Eq.~(\ref{e3.5}) is diagonal in $\alpha,\alpha'$ then the system
corresponds to the excitation of Tomonaga bosons that describe the
electron-hole pairs \cite{schonhammer84}.  Here we seek a useful
approximation to Eq.~(\ref{e3.5}) that describes the excitation of a
system of `average' bosons, the properties of which vary with $z$.
This is achieved by assuming that the dependence of the matrix
elements in Eq.~(\ref{e3.5}) on $z$ and $\alpha,\alpha'$ can be
expressed in the separable form $f_{\alpha,\alpha'}g(z)$.  Physically,
this implies that each electron-hole pair excitation experiences a
time-dependent `force', of the same functional form, but with
different strengths.  Introducing this factorisation leads to
\begin{equation}
P_s(\omega) =\sum_{\sigma}
            \frac{1}{\pi \hbar \omega} \left|
            \int_{-\infty}^{\infty}
            \eta_{\sigma}^{ \frac{1}{2} } (z) \dot{z}(t)
            {\mathrm e}^{-{\mathrm i} \omega t} {\mathrm d}t
            \right|^2
\label{e3.9}
\end{equation}
where different spin terms are shown explicitly (subscript $\sigma$)
and $\eta_\sigma$ is given by Eq.~(\ref{e2.9}) with no sum over spin.
Equation (\ref{e3.9}) is the final expression used to define the FOM
for the calculations reported here.

Although there is no rigorous justification for this simplified form
of $P_s(\omega)$, there are several reasons for expecting it to
provide a useful description of the statistics of the energy
loss. Like Eq.~(\ref{e3.5}), Eq.~(\ref{e3.9}) gives exactly the same
average rate of energy transfer between adsorbate and electron gas as
the classical friction coefficient.  A similar `single channel'
approach has been used by previous authors
\cite{darling95,brako81,gunnarsson82,persson87} in constructing model
descriptions of energy loss processes at surfaces using the FOM, and
this approximation is also exact for other special cases, such as for
symmetry reasons or if the excited electrons are predominately $s$ in
character \cite{schonhammer84}.
In order to implement the FOM in this form only the friction
coefficient and trajectory of the incident particle are required. From
these, the probability distribution function for single electron-hole
pair excitation, $P_s(\omega)$, may be calculated and this function,
through Eq.~(\ref{e3.1}), defines completely the excitation of the
electron gas.

\section{Results}
\label{sec:res}

In this section we present results for the H/Cu(111) adsorbate/surface
system.  A standard Verlet integration of Eq.~(\ref{e2.3}) is
performed with the fixed-spin friction coefficient shown in 
Fig.~\ref{fig2} to obtain the classical trajectory of an incident H atom,
for a range of initial kinetic energies.  The resulting trajectories
show expected features with a critical initial kinetic energy
($\epsilon_c$) above which the atom escapes from the well, and below 
which it is trapped. For H/Cu(111), $\epsilon_c=0.166$ eV, and for an 
atom that escapes from the well we find the round-trip time (for which
$z(t) < 3.0$ \AA) is $\sim 0.04$ ps.

The next step is to apply Eqs.~(\ref{e3.1}), (\ref{e3.2}) and
(\ref{e3.9}) to implement the FOM and extract physically measurable
quantities from it.  A characteristic property of the
adsorbate/surface interaction is the sticking coefficient,
$S(\omega)$, defined as the probability that an incident adsorbate
loses sufficient energy to be captured by the surface.  Here we
consider only the contribution to sticking due to the energy loss via
electron-hole pair production.  In terms of the probability for 
energy gain by the substrate defined
in (\ref{e3.1}) the sticking coefficient becomes
\begin{equation}
S(\omega_i)=\int_{\omega_i}^{\infty} P(\omega) d\omega,
\label{e3.10}
\end{equation}
where $\hbar \omega_i=\epsilon_i$ is the initial kinetic energy of the
adsorbate.  Following the approach described by Sch\"onhammer and
Gunnarsson \cite{schonhammer84} a trajectory is chosen that travels 
into the
surface and out, but is truncated at the next turning point if there
is one.  The use of this trajectory results in a $P(\omega)$ that is
the probability that the electron gas is excited by an energy $\hbar
\omega$ in the time taken for the adsorbate to cross the
surface-adsorbate well twice.  If this energy loss is larger than the
initial kinetic energy of the adsorbate then it is considered to be
captured.  Past applications of the FOM to this type of problem have
generally employed simple model Hamiltonians and elastic trajectories,
hence do not take into account the influence of the loss of energy on
the trajectory itself.  Here calculations have been performed for both
the elastic and truncated inelastic trajectories.

\begin{figure}[t]
\includegraphics[scale=0.95]{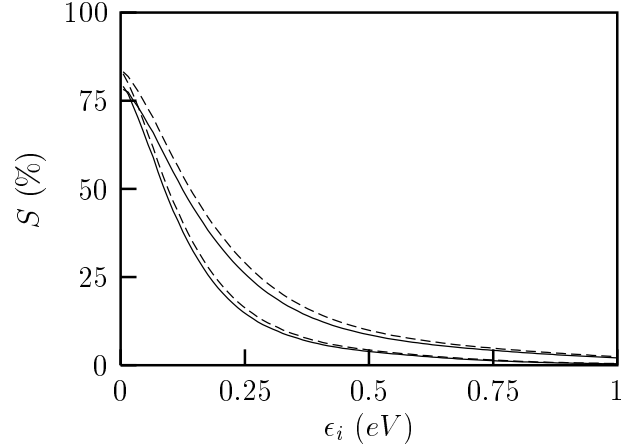}
\caption{\label{fig4}
Sticking coefficients for H and D atoms perpendicularly incident on the top site of Cu(111).
Elastic trajectory is dashed line, inelastic trajectory is solid line. Upper and lower curves
are for H and D respectively.}
\end{figure}

Figure\ \ref{fig4} shows the sticking coefficient as a function of the
incident kinetic energy.  As expected there is an overestimate of
$S(\omega_i)$ for the elastic trajectory due to the increased
velocity.  Generally, $S(\omega_i)$ falls from $\sim 80 \%$ at
$\epsilon_i=0$ to $\sim 40 \%$ at the classical critical initial
kinetic energy, $\epsilon_c$, and then falls smoothly to zero.  It is
important to note that this sticking coefficient does not take into
account any energy loss due to phonon excitations and so it may not be
compared directly to experiment.  This is immediately apparent if we
consider H incident on Cu at $300$K.  Experimental results and
previous theoretical estimates \cite{stromquist98} suggest $S\sim 1$,
whereas for $\epsilon_{i}=\frac{3}{2} kT=38.8$ meV ($T=300$K, the
results vary slowly with incident energy in the thermal range, and
taking a Boltzmann average makes no discernible difference) our
calculation gives $S=0.68-0.72$.

\begin{figure}[b]
\includegraphics[scale=0.95]{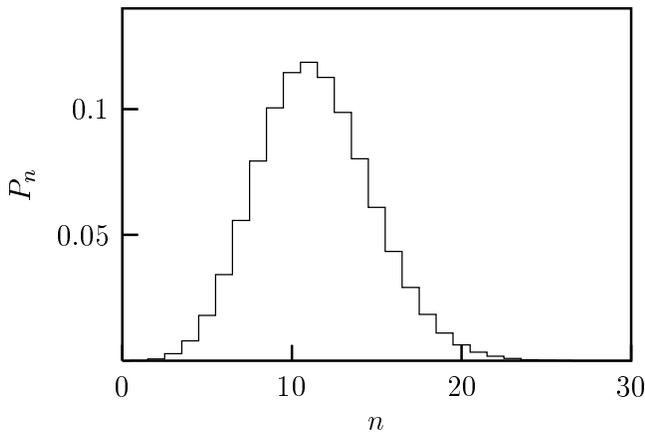}
\caption{\label{fig6}
Probability of excitation process creating $n$ electron-hole pairs, $P_n$.
}
\end{figure}

A quantity of interest is the number of electron-hole pairs excited in
the process of a single H atom being captured by the surface.  The
probability that $n$ electron-hole pairs are excited by the adsorbate
surface interaction is given by the Poisson distribution
\begin{equation}
P_n={\mathrm e}^{-\alpha_0}\alpha_0^n/n!
\label{e3.11}
\end{equation}
where $\alpha_0$ is the average number of electron-hole pairs excited, as
defined earlier.  Figure\ \ref{fig6} shows $P_n$ for H/Cu(111) for an
inelastic trajectory that follows the H atom all the way to the bottom
of the potential well.  $\alpha_0$ takes the value of $11.4$ electrons
which, combined with a total excitation of $\sim 1.8$ eV, implies that
on average each excitation is of order 0.15 eV. This supports the
premise that underlies our approach to calculating the energy transfer
from adsorbate to substrate; namely that the process is nearly
adiabatic and involves the excitation of multiple, low-energy
electron-hole pairs.

\subsection{Chemicurrents}
Of particular interest here, and not investigated before using
\textit{ab initio} methods, is the creation of hot electrons and holes
during the adsorption process.  In a number of papers Nienhaus
\textit{et al.} describe experimental investigations of the electronic
excitation behaviour for a variety of adsorbates and metal surfaces
\cite{nienhaus99,nienhaus00,nienhaus02,gergen01}.  They construct a
Schottky diode device consisting of a clean metal film deposited onto
a Si wafer.  Incident atoms or molecules may excite hot electrons (or
holes) at this metal surface and electrons with energies high enough
to surmount the Schottky barrier can be detected as a current,
referred to as a `chemicurrent'.

The electrons detected in these experiments can be thought of as
undergoing two processes. First they are excited to sufficient energy
to be detected and second a number of geometrical factors and
scattering mechanisms within the detector cause signal loss.  The
second of these has been discussed in detail by Nienhaus
\cite{nienhaus99,nienhaus00,nienhaus02,gergen01} and Gadzuk
\cite{gadzuk02}, and a simple model to describe these processes
results in good agreement between our results and experiment, as
discussed in Ref. \onlinecite{trail02}.  Here we do not consider
device losses, instead we concentrate on the fundamental quantity of
the number of electrons excited above a specific energy. We refer to
this quantity as `electrons made available for detection'.

\begin{figure}[t]
\includegraphics[scale=0.95]{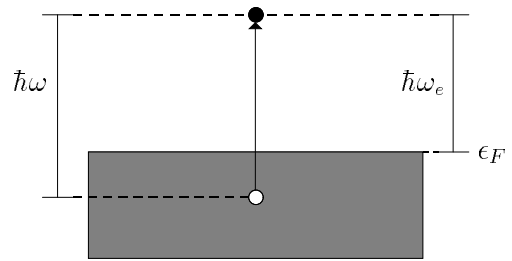}
\caption{\label{fig7} An electron is excited from below the Fermi
energy by $\hbar \omega$ to a state $\hbar \omega_e$ above the Fermi
energy.  The probability for this excitation to occur depends only on
the electron-hole pair energy, $\hbar \omega$, provided $0 < \omega_e
< \omega$.  For $\omega_e > \omega$ and $\omega_e < 0$ this
probability is zero.  }
\end{figure}

The number of electrons made available for detection over a Schottky
barrier of $\hbar \omega_s$ by the adsorbate/surface interaction is
written as $N_e(\omega>\omega_s)$.  To obtain $N_e$ we must transform
from excitation statistics in terms of the energy of electron-hole
pairs to statistics in terms of electron energies.  The probability
that a single excitation event results in an electron-hole pair of
energy $\hbar \omega$, with the electron possessing an energy of
$\hbar \omega_e$, can be written as
\begin{equation}
\frac{P_s(\omega,\omega_e)}{\alpha_0} = \frac{1}{\alpha_0}
  \frac{P_s(\omega)}{\omega} \Theta(\omega-\omega_e) \Theta(\omega_e).
\label{e4.1}
\end{equation}
Here $P_s(\omega)/\alpha_0$ is the probability that a single
excitation event results in an electron-hole pair of energy $\hbar
\omega$, as defined by Eq.~(\ref{e3.9}), and $\Theta(x)=0,1$ for
$x<0,x \ge 0$.  Equation (\ref{e4.1}) is obtained by noting that an
electron-hole pair of energy $\hbar \omega$ will consist of electrons
distributed over the energy range $0 < \hbar \omega_e < \hbar\omega$
with equal probability (see Fig.~\ref{fig7} ).
The probability that a single excitation event results in an electron
of energy $\hbar \omega_e$ may then be obtained by integrating
Eq.~(\ref{e4.1}) over all electron-hole energies, resulting in
\begin{equation}
\frac{P_e(\omega_e)}{\alpha_0}=\frac{1}{\alpha_0}
\int_{\omega_e}^{\infty} \frac{P_s(\omega)}{\omega} d\omega.
\label{e4.2}
\end{equation}
The total number density of electrons excited to energy $\hbar \omega$
by \textit{all} events in a multi-excitation expansion is then
$P_e(\omega)$.  It follows that the number of electrons made 
available for detection can be expressed as
\begin{equation}
N_e(\omega>\omega_s)= \int_{\omega_s}^{\infty} \! {\mathrm d} \omega
                      \int_{\omega }^{\infty} \! {\mathrm d} \omega'
                      \frac{P_s(\omega')}{\omega'}.
\label{e4.3}
\end{equation}

\begin{figure}[t]
\includegraphics[scale=0.95]{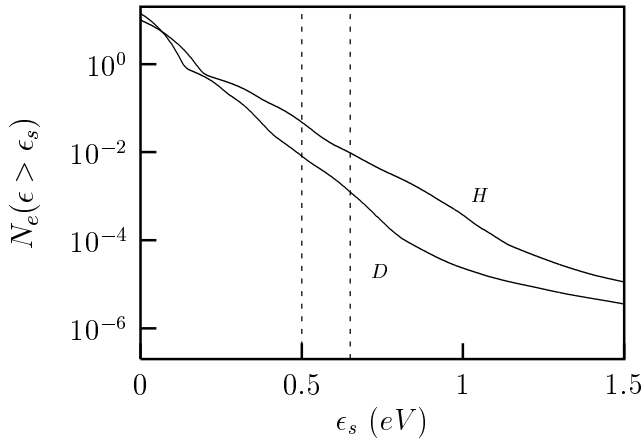}
\caption{\label{fig8} Number of electrons per atom made available for
detection over a Schottky barrier of height $\epsilon_{s}$ for H and D
incident on the top site of Cu(111).  The vertical lines span the
range of Schottky barrier heights found by Nienhaus \textit{et al.}
\cite{nienhaus99} for both Cu and Ag.  }
\end{figure}

\begin{figure}[t]
\includegraphics[scale=0.95]{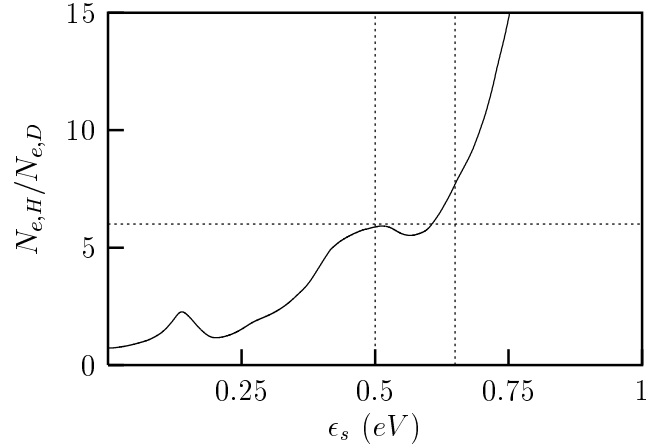}
\caption{\label{fig9} The predicted ratio of the chemicurrents due to
H and D incident on the top site of Cu(111) as a function of Schottky
barrier height.  The vertical lines span the range of Schottky barrier
heights found by Nienhaus \textit{et al.} \cite{nienhaus99} for both
Cu and Ag. The horizontal line is the ratio of chemicurrents found
experimentally for Ag ($\epsilon_{s}=0.5-0.55$ eV).  }
\end{figure}

Figure\ \ref{fig8} shows $N_e(\omega>\omega_s)$ for a range of barrier
heights, and for H and D normally incident on the top site of Cu(111).
As in Fig.~\ref{fig4} the incident energy is 38.8 meV, and full
inelastic trajectories are used.  The data shows an essentially
exponential behaviour over most of the range considered, and most of
the difference between the results for H and D may be accounted for by
scaling $\epsilon_s$ by $\sqrt{2}$ (the agreement would be exact for
an elastic trajectory).  This dynamic scaling is a consequence of
Eq.~(\ref{e3.9}) and the scaling of velocity with mass for a given
kinetic energy.  The unusually strong isotope effect apparent in
Fig.~\ref{fig8} is consistent with the experimentally measured
chemicurrent.  Nienhaus \textit{et al.} \cite{nienhaus99} reported no
results for D incident on Cu, but for Ag the detected number of
electrons per atom for H was larger than for D by a factor of $\sim
6$.  Figure\ \ref{fig9} shows the calculated ratio of chemicurrents
for H and D, and in the range of relevant barrier heights we find an
isotope effect close to the experimental result.  This comparison is
particularly useful as it is not necessary to consider device losses
if we take the signal to be suppressed by the same factor for both H
and D incident upon the detector.

It is interesting to ask where in the trajectory most of the 
relatively high-energy electrons that are detected as a chemicurrent
are produced. This is significant because of the discussion in 
Section I concerning our method for removing the singularity in the 
friction coefficient caused by the spin transition.  By keeping 
the spin fixed the singularity is removed, but the friction coefficient 
is still peaked at $z_0=2.39$\ \AA. Given the rather ad-hoc way 
in which this peak is derived, if we find that it makes a major 
contribution to the predicted chemicurrent then the basis of our 
results could be questioned. If, however, the chemicurrent is relatively 
insensitive to the peak in $\eta$, then we can have more confidence in 
our results. Equation (\ref{e3.9}) shows that the spectrum of 
electron-hole pairs is determined by two factors, the speed of the
adsorbate, $\dot{z}$, and the friction coefficient, $\eta$. These 
peak in very different places in the trajectory; $\dot{z}$ is 
greatest at the minimum of the potential well (at about 1.4\ \AA) 
while $\eta$ peaks at $z_0=2.39$\ \AA. Which dominates the chemicurrent?

\begin{figure}[b]
\includegraphics[scale=0.95]{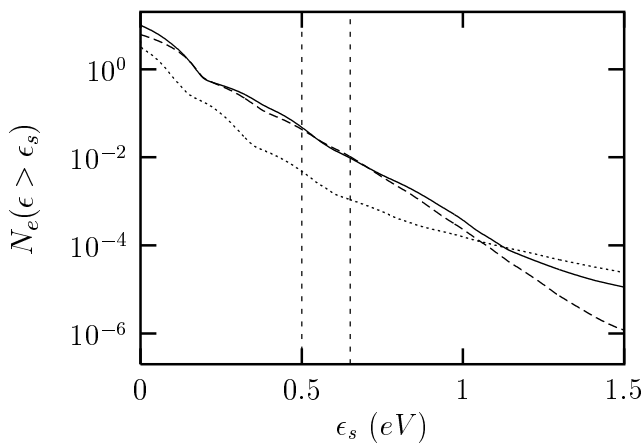}
\caption{\label{fig8b}
Number of electrons per atom made available for detection over a
Schottky barrier of height $\epsilon_{s}$ for H incident on the top
site of Cu(111).  The vertical lines span the range of Schottky
barrier heights found by Nienhaus \textit{et al.} \cite{nienhaus99}
for both Cu and Ag.  Solid line is $N_e$ due to the full friction
coefficient (as in Fig.~\ref{fig2}), dashed line that due $\eta^{sm}$,
the smooth part of the profile, and the dotted line that due to
$\eta^{pk}$, the peaked part of the profile.
}
\end{figure}

To address this question we (arbitrarily) separate the friction 
coefficient into two parts, a `smooth' part

\begin{equation}
\eta_{\sigma}^{sm}(z)=\left\{
   \begin{array}{ll}
   \eta_{\sigma}(z)                                & z \leq 1.85\ {\textnormal \AA} \\
   \eta_{\sigma}(1.85) {\mathrm e}^{-2(z-1.85)^2}  & z >    1.85\ {\textnormal \AA}
   \end{array}
\right.
\label{e4.4}
\end{equation}
and a `peak' part that is zero for $z \leq 1.85\ {\textnormal \AA}$,
\begin{equation}
\eta^{pk}_{\sigma}(z)=\eta_{\sigma}(z)-\eta_{\sigma}^{sm}(z).
\label{e4.5}
\end{equation}
The value $z=1.85\ {\textnormal \AA}$ is chosen to be the minimum of
$\eta$ between $z=1$ and $2\ {\textnormal \AA}$ (see Fig.~\ref{fig2}),
and the function introduced in Eq.~(\ref{e4.4}) is chosen to decay
smoothly from $z=1.85\ {\textnormal \AA}$ to a value below the total
friction at $z=4\ {\textnormal \AA}$.  This continuous decay prevents
a discontinuity in the integrand of Eq.~(\ref{e3.9}) which would
result in spurious high energy electron-hole pair excitations.
Calculations are carried out as before, using the full friction
coefficient to obtain the trajectories, but using either
$\eta_{\sigma}^{sm}$ or $\eta_{\sigma}^{pk}$ to evaluate measurable
quantities from the FOM.  This allows us to analyse the statistics of
electron-hole pair creation due to each part of the friction
coefficient separately.  For example, we find that $\sim 25 \%$ of the
total energy transfered from the incident atom to electron-hole pairs
is due to the peak, indicating that this region is significant, but
not of primary importance.  Figure\ \ref{fig8b} shows
$N_e(\epsilon>\epsilon_s)$ obtained from the smooth and peaked parts,
in comparison to that obtained from the complete friction profile.  It
should be stressed that the sum of the first two quantities is not
equal to the last, due to the factor of $\sqrt{\eta}$ in
Eq.~(\ref{e3.9}). It can be seen that the presence or absence of the
peaked region of $\eta$ makes little difference to the predicted
number of electrons made available for detection, at least for the
Schottky barrier heights considered here.  We conclude that the
interaction of the incident atom with the electron-hole pairs of the
substrate is not dominated by the peak found in the calculated
friction coefficient.  This suggests that, provided the peak does not
become singular, its precise form will not have a great bearing on our
predictions of experimental observables.

Finally, we have not so far addressed an important question: what
influence would the introduction of energy loss via the creation of
phonons have on our estimates for the detected chemicurrent?  A full
answer to this would require a molecular dynamics description of the
interaction, including electron-hole pair creation, and is beyond
currently available computing resources.  However, Str\"omquist
\textit{et al.} \cite{stromquist98} have performed molecular dynamics
calculations for H/Cu(111) (not including electron-hole pair creation)
using a model energy surface fitted to \textit{ab initio} data,
and sampling a large number of trajectories.  They obtain an energy
relaxation rate of $\sim 0.7$ ps$^{-1}$ due to phonon creation, a
value that is similar to the rate of energy loss due to electron-hole
pairs for the perpendicular vibrational mode of H on Cu(111) discussed
above.  We may obtain a crude estimate for the influence of phonon
creation on our results by introducing an `extra' friction term to the
calculation of the trajectory (not the FOM itself) that would
reproduce this energy loss rate were no electron-hole pair creation
present.  Taking an appropriate friction coefficient and introducing
the trajectory at $z=4$ \AA\ above the surface (see Fig.~\ref{fig2})
results in a decrease in the chemicurrent by an approximately constant
factor of $\sim 0.8$ for H, and a decrease in the average number of
excited electron-hole pairs from $11.4$ to $9.5$.  Although we stress
this is a crude estimate it does suggest that phonon production has a
small (though not negligible) influence.  For heavier adsorbates a
stronger effect would be expected.

\section{Conclusion}
An \textit{ab initio} description of the energy transfer from an
adsorbate incident on a metal surface to the electrons present in the
surface via electron-hole pair creation has been developed.  Energy
loss from the adsorbate has been described semi-classically, and the
excitation of the electron gas described quantum mechanically using a
FOM.  It has been shown that a large degree of consistency exists
between TDDFT, the classical energy loss of the adsorbate and the
energy gain of the substrate electrons, in that the average
energy changes are equal.  We have also shown that the Markov
limit used to define a friction coefficient for the classical motion of 
the adsorbate has a counterpart in a common Fermi energy approximation 
for the matrix elements used in the implementation of the FOM.

This classical adsorbate/quantum electronic description of the
non-adiabatic electron-hole excitation process has been applied to H
(and D) incident on Cu(111), with somewhat surprising results.  We
find that a singular friction coefficient results from the application
of spin-dependent DFT.  The singular behaviour occurs at the
transition point (spin-polarised to spin-degenerate) and is due to the
strongly non-adiabatic nature of the evolution of the system around
this transition.  A breakdown of the nearly-adiabatic definition of the
friction coefficient results, suggesting that the friction coefficient
cannot be defined for systems exhibiting a spin-transition of this
kind.  The FOM exhibits the same singular behaviour.  Further
investigation of this effect suggests that the strongly non-adiabatic
system can be replaced by a weakly non-adiabatic `fixed spin' system
that provides a good approximation for the trajectories of the
incident atoms considered, avoiding the breakdown of both the FOM and
the friction description.  Our final results indicate that, provided
the singularity in the friction coefficient is removed, the region of
the trajectory in the vicinity of the spin transition is not of
prime importance.

Comparison of the results presented here with the `chemicurrent'
detected by Nienhaus \textit{et al.} \cite{nienhaus99} using Schottky
diode devices has been given in a previous Letter \cite{trail02}.
Calculated and experimental chemicurrents agree well, and we have
shown that simple dynamics can reproduce the large difference between
measured chemicurrents for H and D.  The calculations presented here
can be extended in several ways. For an H atom adsorbate it would be
interesting to investigate other sites on the surface and also
to consider
motion parallel to the surface. Other atomic adsorbates and surfaces
can obviously also be analysed, although for heavier species it will
be important to treat phonon as well as electron-hole pair
excitation. It would also be interesting to consider molecular
adsorbates. One question that immediately arises here is whether
vibrational or rotational motion is as effective as translation in
producing electron-hole pairs.

\begin{acknowledgements}
We acknowledge the support of the UK Engineering and Physical Sciences 
Research Council (EPSRC) and M.P. is grateful for support from the 
Swedish Research Council (VR) and the Swedish foundation for strategic 
research (SSF) through the materials consortium `ATOMICS'.
\end{acknowledgements}

\appendix*

\section{}
\label{sec:appa}

In this appendix we outline the derivation of the results in
Eqs.~(\ref{eq:nonad}) and (\ref{e3.6}). To ease the notation 
we take $\hbar=1$.
We begin by deriving the general expression for the non-adiabatic
energy transfer in Eq.~(\ref{eq:nonad}),  which is just a variant of
the Hellmann-Feynmann theorem. The instantaneous expectation value
$E(t)$ of the many-electron system in the presence of the adsorbate is
defined as
\begin{equation} 
E(t) = \langle \psi(t)|\hat{H}_0 + \hat{V}_{\textrm{ext}}(t)|\psi(t)\rangle
\label{eq:EDef}
\end{equation}
where $\hat{H}_0$ is the many-electron Hamiltonian, $\hat{V}_{\rm
ext}(t)$ is the time-dependent interaction of the electrons with the
moving nucleus, and $|\psi(t)\rangle$ is the many-electron state 
satisfying the time-dependent Schr{\"o}dinger equation with the 
Hamiltonian
$\hat{H}(t)=\hat{H}_0 + \hat{V}_{\textrm{ext}}(t)$. Using the
Schr{\"o}dinger equation in the evaluation of the time-derivative of
$E(t)$ one obtains directly
\begin{widetext}
\begin{eqnarray}
\dot{E}(t) & = & -i\langle \psi(t)|\hat{H}(t)\hat{H}(t)|\psi(t)\rangle + 
              i\langle \psi(t)|\hat{H}(t)\hat{H}(t)|\psi(t)\rangle + 
              \langle \psi(t)|\frac{\hat{H}(t)}{dt}|\psi(t)\rangle \\
           & = & \langle \psi(t)|\frac{\hat{V}_{\textrm{ext}}(t)}{dt}|\psi(t)\rangle.
\label{eq:DotERes}
\end{eqnarray}
To obtain the non-adiabatic part of $\dot{E}(t)$ we have to subtract
the adiabatic part $\dot{E}_0(t)$. The adiabatic energy ${E}_0(t)$ is
defined in an analogous manner to ${E}(t)$ in Eq.~(\ref{eq:EDef}) but
with the difference that $|\psi(t)\rangle = |\psi_0(t)\rangle$ where
$|\psi_0(t)\rangle$ is the instantaneous ground state of $\hat{H}(t)$,
that is, $\hat{H}(t)|\psi_0(t)\rangle =
E_0(t)|\psi_0(t)\rangle$. Using this definition of
$|\psi_0(t)\rangle$, a straightforward differentiation gives
\begin{eqnarray}
\dot{E}_0(t) & = & \langle \dot{\psi}_0(t)|E_0(t)|\psi_0(t)\rangle + 
              \langle \psi_0(t)|E_0(t)|\dot{\psi}_0(t)\rangle + 
              \langle \psi_0(t)|\frac{\hat{H}(t)}{dt}|\psi_0(t)\rangle \\
 & = & \langle \psi_0(t)|\frac{\hat{V}_{\textrm{ext}}(t)}{dt}|\psi_0(t)\rangle \ .
\label{eq:DotE0Res}
\end{eqnarray}
\end{widetext}
In the last step we have used the fact that the derivative of the norm
of $|\psi_0(t)\rangle$ is zero, that is, $\frac{d}{dt}\langle
\psi_0(t)|\psi_0(t)\rangle = \langle \dot{\psi}_0(t)|\psi_0(t)\rangle
+ \langle \psi_0(t)|\dot{\psi}_0(t)\rangle = 0$. Eq.~(\ref{eq:nonad}) is
now directly proved by using the explicit result for $\hat{V}_{\rm
ext}(t) = \int d\mathbf{r} V_{\rm
ext}(\mathbf{r},z(t))\hat{n}(\mathbf{r})$ where $V_{\rm
ext}(\mathbf{r},z)$ is the bare electron-nucleus interaction potential
and $\hat{n}(\mathbf{r})$ is the electron density operator. Note that
$n(\mathbf{r},t) \equiv \langle
\psi(t)|\hat{n}(\mathbf{r})|\psi(t)\rangle$ and $n_0(\mathbf{r},t)
\equiv \langle \psi_0(t)|\hat{n}(\mathbf{r})|\psi_0(t)\rangle$.

We now turn to the derivation of the expression for the non-adiabatic
energy transfer in Eq.~(\ref{e3.6}).  Starting with $P_t(\omega)$, the
probability of the incident atom having lost energy $\omega$ by
electron-hole excitation at time $t$, the expectation value of the
energy loss is given by
\begin{equation}
E_t=\int_{-\infty}^{\infty} \omega P_t(\omega) d\omega
\label{eA.1}
\end{equation}
where
\begin{equation}
P_t(\omega)=\frac{1}{2 \pi} \int_{-\infty}^{\infty}
            {\mathrm e}^{P_{s}(\tau,t)-\alpha_0(t)}
          {\mathrm e}^{ {\mathrm i} \omega \tau} d \tau
\label{eA.2}
\end{equation}
and $P_{s}(\omega,t)/\alpha_0(t)$ is defined as in Eq.~(\ref{e3.4}), but
with the upper limit of the time integral taken as $t$.  Physically
this quantity is the probability that a single electron-hole
excitation event is of energy $\omega$, in the time interval $-\infty$
to $t$.  Interpreting the factor $i\omega$ as a derivative operator in
the time co-ordinate leads to
\begin{eqnarray}
E_t & = &  \left.
      i \frac{d}{d\tau} {\mathrm e}^{P_{s}(\tau,t)-\alpha_0(t)}
      \right|_{\tau=0}  \\ \nonumber
    & = &  \left.
      i \frac{d}{d\tau} P_{s}(\tau,t) \right|_{\tau=0}.
\label{eA.3}
\end{eqnarray}
$P_{s}(\tau,t)$ is then expressed in terms of its Fourier transform 
in $\tau$, giving 
\begin{equation}
E_t=\int_{-\infty}^{\infty} \omega P_s(\omega,t) d\omega.
\label{eA.4}
\end{equation}
It is also possible to obtain higher moments of $P_t(\omega)$ in terms
of $P_s(\omega,t)$ by the same approach.



\begin{thebibliography}{}
\bibitem{fixednote}     In calculations, the spin is fixed by keeping separate Fermi energies for the spin up and spin down electrons.
\end{thebibliography}
\end{document}